# Hard ferromagnetism down to the thinnest limit of iron-intercalated tantalum disulfide

Samra Husremović,[a] Catherine K. Groschner,[a] Katherine Inzani,[b,c,†] Isaac M. Craig,[a] Karen C. Bustillo,[d] Peter Ercius,[d] Nathanael P. Kazmierczak,[a,‡] Jacob Syndikus,[a] Madeline Van Winkle,[a] Shaul Aloni,[c] Takashi Taniguchi,[e] Kenji Watanabe,[f] Sinéad M. Griffin,[b,c] D. Kwabena Bediako[a,g]*

[a] *Department of Chemistry, University of California, Berkeley, California 94720, United States*

[b] *Materials Sciences Division, Lawrence Berkeley National Laboratory, Berkeley, CA 94720, United States*

[c] *The Molecular Foundry, Lawrence Berkeley National Laboratory, Berkeley, California 94720, United States*

[d] *National Center for Electron Microscopy, Molecular Foundry, Lawrence Berkeley National Laboratory, Berkeley, California 94720, United States*

[e] *Research Center for Functional Materials, National Institute for Materials Science, Tsukuba 305-0044, Japan*

[f] *International Center for Materials Nanoarchitectonics, National Institute for Materials Science, Tsukuba 305-0044, Japan*

[g] *Chemical Sciences Division, Lawrence Berkeley National Laboratory, Berkeley, California 94720, United States*

[†] *Current position: School of Chemistry, University of Nottingham, Nottingham, NG7 2RD, United Kingdom*

[‡] *Current position: Division of Chemistry and Chemical Engineering, California Institute of Technology, Pasadena, CA 91125, United States*

* *E-mail:* bediako@berkeley.edu

**ABSTRACT**

Two-dimensional (2D) magnetic crystals hold promise for miniaturized and ultralow power electronic devices that exploit spin manipulation. In these materials, large, controllable magnetocrystalline anisotropy is a prerequisite for the stabilization and manipulation of long-range magnetic order. In known 2D magnetic crystals, relatively weak magnetocrystalline anisotropy results in typically soft ferromagnetism. Here, we demonstrate that ferromagnetic order persists down to the thinnest limit of $Fe_xTaS_2$ (Fe-intercalated bilayer 2*H*-$TaS_2$) with giant coercivities up to 3 tesla. We prepare Fe-intercalated $TaS_2$ by chemical intercalation of van der Waals layered 2*H*-$TaS_2$ crystals and perform variable-temperature quantum transport, transmission electron microscopy, and confocal Raman spectroscopy measurements to shed new light on the coupled effects of dimensionality, degree of intercalation, and intercalant order/disorder on the hard ferromagnetic behavior of $Fe_xTaS_2$. More generally, we show that chemical intercalation gives access to a rich synthetic parameter space for low-dimensional magnets, in which magnetic properties can be tailored by the choice of the host material and intercalant identity/amount, in addition to the manifold distinctive degrees of freedom available in atomically thin, van der Waals crystals.

## INTRODUCTION

The recent discovery of long-range magnetic ordering in the atomically thin layers of $Cr_2Ge_2Te_6$,[1] $CrI_3$,[2] and $Fe_3GeTe_2$[3] has spurred intense interest into the development of "two-dimensional" (2D) magnetic crystals. The experimental paradigm that has generated the most consistent success in this area has generally been to exfoliate layered van der Waals (vdW) crystals that are known to exhibit magnetic ordering in the bulk. In addition to serving as promising candidates for a new generation of compact, energy-efficient spin-based electronics,[4–7] these 2D magnets are exceptional platforms for exploring fundamental principles pertaining to the stabilization of long-range magnetic order at low dimensions. According to the Mermin–Wagner–Hohenberg theorem, long-range magnetic order in 2D systems should not be stable in an isotropic model.[8,9] However, intrinsic magnetocrystalline anisotropy (MCA)—which imparts a preferential spin orientation—plays an integral role in stabilizing the long-range magnetic order in 2D vdW magnets.[10,11] Still, the overwhelming majority of ferromagnetic 2D crystals are soft, possessing weak coercive fields, which arise from relatively weak MCA.[12] The development of new atomically thin materials with strong and predictable control over MCA is imperative for tailoring magnetism in 2D materials that may serve as testbeds for fundamental theory and pave the way towards ultralow power spintronic technologies.

Transition metal dichalcogenides (TMDs) intercalated with open-shell transition metals comprise a family of materials with inherently tunable MCA, which makes them appealing targets for realizing 2D magnets with deterministic properties. The strength of MCA in intercalated TMDs is determined by the respective magnitudes of spin-orbit coupling (SOC) and unquenched orbital angular momentum (OAM) of the spin-bearing ions.[13–16] While the SOC is increased when the host lattice or the intercalants comprise heavy elements, the OAM of spin-bearing ions is shaped by their oxidation states and coordination environments imposed by the interlayer galleries.[13–18] However, low-dimensional analogues of these systems have not been magnetically characterized; strong interlayer interactions in the effectively ionic, bulk crystals of these intercalation compounds have, so far, precluded exfoliation of pristine, large 2D flakes suited for magnetic studies.[19–22] Consequently, the manifold possibilities for designing a new class of 2D magnets with controllable MCA by exploiting the tunable vdW interface, host lattice, and intercalant identity/stoichiometry remain untapped.

In this work, we discover the persistence of long-range ferromagnetism in a transition metal-intercalated TMD, $Fe_xTaS_2$, down to the 2D limit. We show that chemical intercalation of iron between layers of the exfoliated, pristine host lattice $2H$-$TaS_2$ gives access to this intercalation compound with controlled dimensionality from few layers down to the thinnest limit, Fe-intercalated bilayer $2H$-$TaS_2$. We find very strong MCA in these 2D materials, associated with the high unquenched OAM of the trigonal antiprismatic (*i.e.* trigonally elongated pseudo-octahedral) Fe centers and SOC from the Ta-based host. This large MCA is reflected in magnetotransport measurements, which reveal that anisotropy-stabilized hard ferromagnetic behavior is retained in the 2D limit, producing giant coercive fields up to 3 T. Furthermore, we find that the degree of intercalation and order/disorder in intercalated species heavily influences the magnetoelectronic behavior, establishing this family of intercalation compounds as versatile platforms for designing 2D magnets.

## RESULTS AND DISCUSSION

### Fabrication of vdW heterostructures, chemical intercalation, and structural characterization

Few-layer $Fe_xTaS_2$ was synthesized by intercalating iron into mechanically exfoliated $2H$-$TaS_2$ flakes using a procedure adapted from established methods for soft chemical intercalation of metals using molecular precursors containing zerovalent metal centers.[23–27] In a typical procedure, $2H$-$TaS_2$ flakes were mechanically exfoliated onto $SiO_2$/Si substrates and capped with electrically insulating hexagonal boron nitride (hBN) to prevent degradation of the $TaS_2$ basal plane,[28] while exposing one or more edges of the TMD flake. Under inert atmosphere, the $SiO_2$/Si substrates bearing these vdW heterostructures are immersed in a 10 mM solution of $Fe(CO)_5$ in toluene, acetone, or acetonitrile at 50 – 55 °C for about 24 hours as depicted in **Figure 1a** (See SI for experimental details). After washing with fresh solvent, samples were annealed under high vacuum ($10^{-7}$ torr) at 350 °C for 30 minutes to promote ordering of intercalants (SI Figure 5).[24]

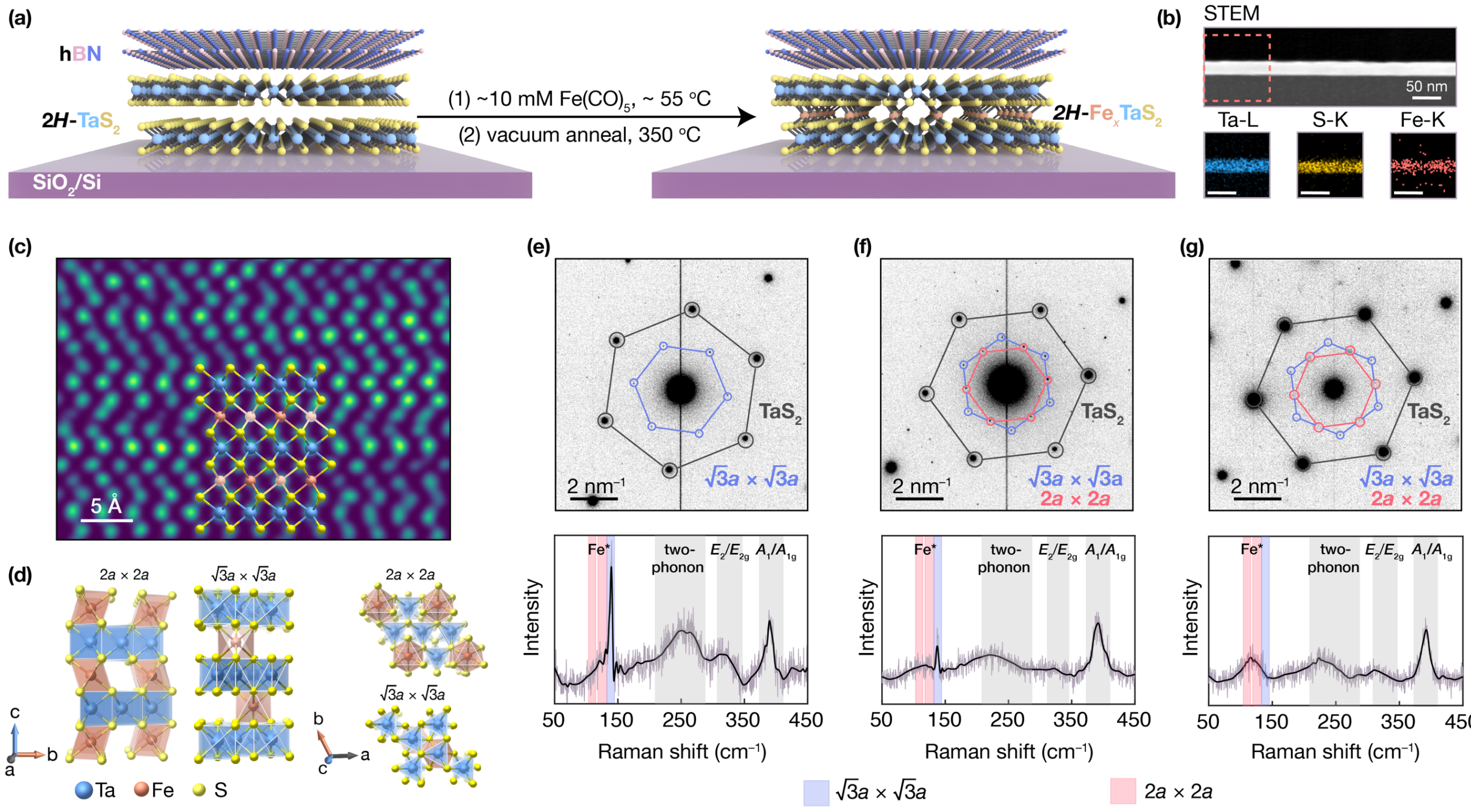

**Figure 1. (a)** Schematic illustration of iron intercalation into the hBN/2*H*-$TaS_2$ heterostructure and formation of $Fe_xTaS_2$. **(b)** STEM-EDS scans perpendicular to the *c*-axis of a cross-sectional $Fe_xTaS_2$ sample (white region in STEM image). The dashed red box in the STEM image marks the area where EDS data was collected. **(c)** DPC-STEM micrograph of a cross-sectioned $Fe_xTaS_2$ sample along the $[10\bar{1}0]$ zone axis of the 2*H*-$TaS_2$ lattice. The structure of $Fe_{1/3}TaS_2$ (ref 30) is overlaid with the micrograph. **(d)** Published crystal structures for bulk $Fe_{0.25}TaS_2$ (ref 29) and $Fe_{0.33}TaS_2$ (ref 30). **(e)–(g)** Pairs of Raman spectra and NBED patterns of three chemically intercalated $Fe_xTaS_2$ flakes on electron-transparent $Si_3N_4$ grids. In each case, Raman spectra and NBED data were measured in identical regions of the flake. The first order Bragg spots of the 2*H*-$TaS_2$ host lattice and the Fe superlattices are marked in the NBED patterns. Raman spectra are normalized to the $A_1/A_{1g}$ peak, and the spectral regions where phonon modes for different superlattices are expected[32] are highlighted. The raw data are represented in light gray, while dark gray lines represent denoised traces.

To examine the composition and structure of our $Fe_xTaS_2$ thin crystals, we obtained and high-resolution TEM (SI Figure 6), scanning TEM (STEM) combined with energy-dispersive X-ray spectroscopy (EDS) and differential phase contrast (DPC)-STEM with atomic resolution (See SI for experimental details). The *c*-axis HRTEM (SI Figure 6) data verify that the *ab* atomic periodicities match the expected values for bulk $Fe_xTaS_2$, and EDS measurements of cross-sectioned thin crystals confirm the incorporation of Fe into $TaS_2$ after chemical intercalation (**Figure 1b**). The Fe intercalants occupy 6-coordinate interstitial sites between the layers of 2*H*-$TaS_2$, as revealed by DPC-STEM of cross-sectioned $Fe_xTaS_2$ samples (**Figure 1c,** SI Figure 8, 10) displaying a prominent $\sqrt{3}a \times \sqrt{3}a$ superstructure in their Raman spectra (SI Figure 9). Further analysis of our STEM-DPC micrographs (See SI for details) shows that the pseudo-octahedral Fe coordination environments are trigonally elongated by 8.0(6) %. The magnitude of this experimentally determined distortion is congruous with the ~8 % trigonal distortion predicted by our DFT calculations. The trigonally elongated pseudo-octahedral Fe coordination has also been reported for bulk $Fe_xTaS_2$ prepared via high-temperature chemical vapor transport crystal growth (**Figure 1d**).[29,30]

In bulk $Fe_xTaS_2$ analogs, depending on the stoichiometry, superlattices with $2a \times 2a$ or $\sqrt{3}a \times \sqrt{3}a$ periodicity are observed (where $a$ = the lattice constant of 2*H*-$TaS_2$), with perfect superlattices at $x = 0.25$ and 0.33, respectively (**Figure 1d**).[29–31] To examine if analogous intercalant ordering is present in our chemically treated few-layer materials, we measured both nanobeam electron diffraction (NBED) patterns and Raman spectra within 500 nm of each other on samples mounted on silicon nitride transmission electron microscopy (TEM) grids (See SI for experimental details). As shown in **Figures 1e–g**, NBED patterns of $Fe_xTaS_2$ samples prepared by chemical intercalation and annealing exhibit superlattice spots that are not present in 2*H*-$TaS_2$ (SI Figure 5), and the accompanying Raman spectra display a new set of peaks in the low-frequency region (100 – 150 $cm^{-1}$). These fingerprints of Fe ordering are consistent with previous literature on bulk $Fe_xTaS_2$.[32,33] For example, the sample that displays sharp $\sqrt{3}a \times \sqrt{3}a$ superlattice spots in NBED data also exhibits an intense Fe-in-plane (Fe*) mode at 140 $cm^{-1}$, which is characteristic of the $\sqrt{3}a \times \sqrt{3}a$ superstructure (**Figure 1e).**[32] In contrast, as shown in **Figures 1f,g**, samples that evince two clear superstructures (both $\sqrt{3}a \times \sqrt{3}a$ and $2a \times 2a$) in NBED data are characterized by generally broader and less intense Raman features that span the energy ranges expected for each type of Fe superlattice ordering. For these dual-superlattice samples, sharp Fe-related Raman features may be absent (**Figure 1g**) or show a red-shift (**Figure 1f**) compared to the crystals that host a single, dominant Fe superstructure (**Figure 1e**). Further, a comparison of these data reveals that the sharpness of Fe-related NBED spots and Raman-active Fe* phonon modes appear to be correlated; samples with diffuse superlattice spots in NBED show broad Fe* features in Raman spectra (**Figure 1g**), while samples with sharp superstructure spots in NBED present more well defined Fe* peaks in Raman spectra (**Figure 1e,f**). These data demonstrate that chemical intercalation is a viable method for accessing $Fe_xTaS_2$ in the few-layer limit with the potential to vary stoichiometry and homogeneity. We find that all Raman spectra in **Figure 1e–g** display a red shift of the $A_{1g}$ mode by 8–10 $cm^{-1}$ from the expected value for 2*H*-$TaS_2$.[34,35] This shift is consistent with electron transfer from Fe to the host lattice,[36,37] in agreement with our DFT calculations. For all calculated structures and magnetic configurations, we found that Fe converges to a +2 oxidation state, matching our electron energy loss spectroscopy (EELS) results (SI Figure 12) and published experimental data on bulk $Fe_xTaS_2$.[38–40] We note that without annealing these chemically intercalated materials, we neither observe superlattice spots in NBED data nor significant changes to the Raman spectra (SI Figure 5). These observations suggest that a modest thermal treatment is required both for ordering of Fe in the vdW interface between $TaS_2$ layers[24] as well as charge transfer from the intercalants to the host lattice (see also SI Section 2.2).

**Variable temperature transport measurements**

We interrogated the magnetoelectronic behavior of chemically intercalated $Fe_xTaS_2$ flakes by fabricating mesoscopic Hall bar devices for variable temperature magnetotransport measurements (**Figure 2a**) (See SI for details on device fabrication). In these experiments, we simultaneously monitored the longitudinal ($\rho_{xx}$) and transverse Hall resistivity ($\rho_{xy}$) as a function of temperature ($T$) and external magnetic field ($H$) to evaluate magnetic order in our materials via its effect on electronic transport. **Figure 2** presents a representative set of results for a hBN/11-layer $Fe_xTaS_2$ heterostructure on a supporting $SiO_2$/Si substrate (See SI for details on determining the layer count). Confocal Raman spectroscopy data (**Figure 2b**) acquired along the measurement channel of the Hall bar are consistent with a dominant $\sqrt{3}a \times \sqrt{3}a$ superstructure that gives rise to a strong peak at 139 $cm^{-1}$, albeit showing evidence for the co-existence of a much less dominant $2a \times 2a$ superlattice in a weaker broad feature spanning 100–135 $cm^{-1}$. The co-existence of two superlattices can be a result of either stoichiometric inhomogeneities within the spot size (*ca*. 500 nm) of the Raman laser and/or deviations of $x$ from the commensurate amounts of 1/4 and 1/3.[20,41] These possibilities were evaluated by cross-sectioning devices with focused ion beam and using STEM-EDS to map the Fe–Ta ratio (see SI for experimental details). These measurements reveal an effectively consistent Fe:Ta value of 0.26 ± 0.02 along the Hall bar channel (**Figure 2c**).

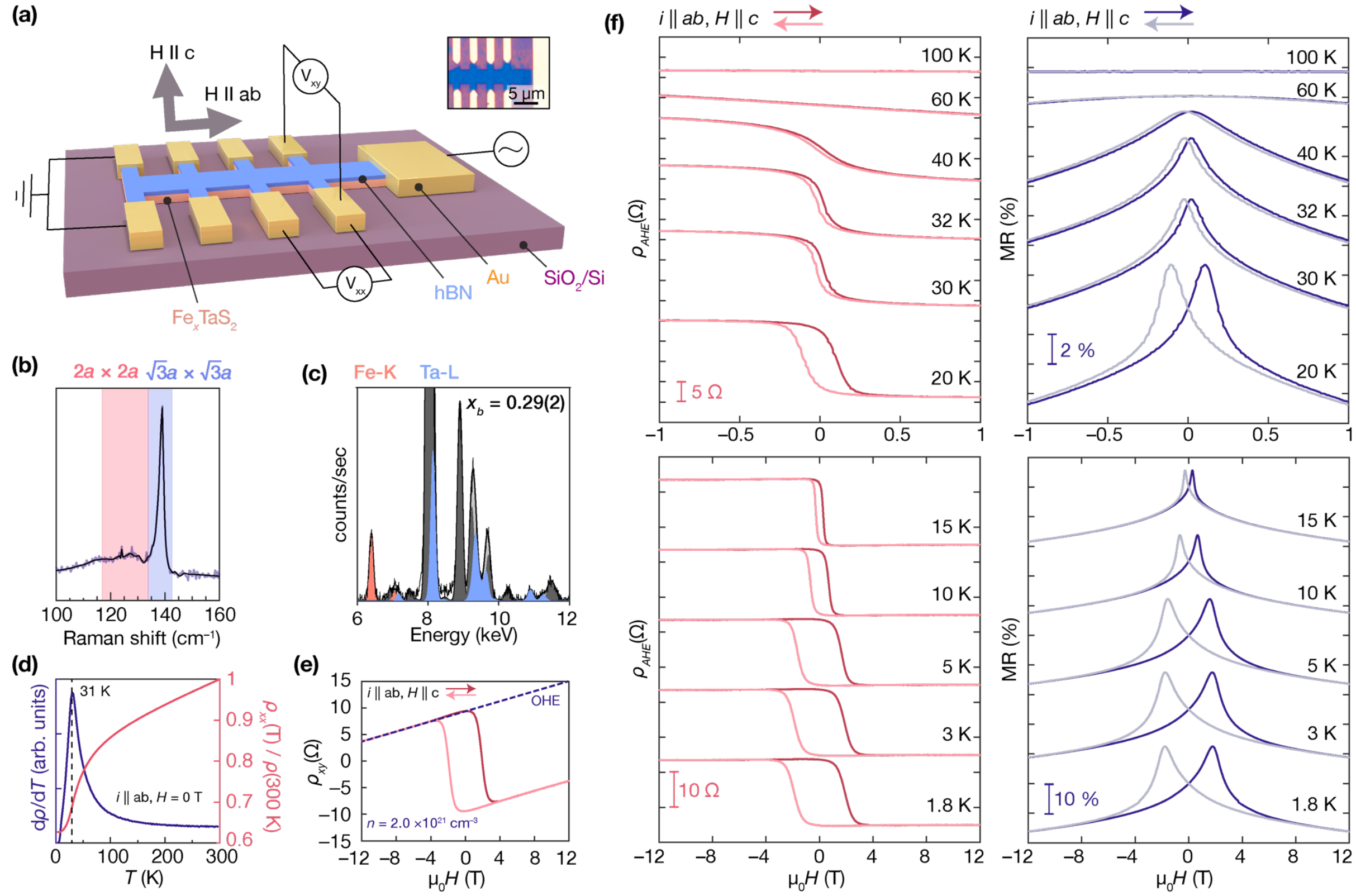

**Figure 2. (a)** Schematic of the hBN/11-layer $Fe_xTaS_2$ device on $SiO_2$/Si. **(b)** Cumulative low-frequency Raman spectrum across the Hall bar channel of the device. **(c)** Representative cross-sectional STEM-EDS of the device. **(d)** Temperature-dependent longitudinal resistivity and its first derivative with respect to temperature of the hBN/11-layer $Fe_xTaS_2$ heterostructure. **(e)** Hall resistivity at 1.8 K with the ordinary Hall effect (OHE) highlighted. **(f)** Temperature dependence of anomalous Hall resistivity $\rho_{AHE}$ (left), and magnetoresistance, MR (right), as a function of $H$.

To compare the Fe:Ta stoichiometry measured in thin flakes with the values determined for bulk (3D) crystals, we must consider that terminating Ta ions from the top and bottom $TaS_2$ layers contribute substantially to the measured elemental composition of thin crystals. Consequently, the ratio of Fe to Ta should be lower in thin samples compared to their structurally equivalent 3D analogues. We can convert between the Fe:Ta values measured in thin samples ($x_t$) possessing $L$ number of layers and the stoichiometry of their compositionally equivalent bulk analogues ($x_b$) using the following expression (see SI for details on the simple derivation):

$$x_b = x_t / \left(1 - \frac{1}{L}\right) \quad (1)$$

With this expression, we find that the equivalent bulk stoichiometry, $x_b = 0.29 \pm 0.02$ for our sample shown in **Figure 2**, confirming that the degree of intercalation represents a slight deviation from the commensurate value of $x$ = 1/3. To facilitate comparison with the magnetotransport behavior of bulk crystals, all stoichiometries given for thin samples in this paper are converted to $x_b$.

Temperature-dependent longitudinal resistance measurements revealed a phase transition upon cooling. An inflection in $\rho_{xx}$ and corresponding peak in $d\rho_{xx}/dT$ at 31 K indicate that the sample underwent a phase transition at this temperature (**Figure 2d**).[42] To probe the nature of this low-temperature phase, we measured the field-dependent Hall resistivity, $\rho_{xy}$, at 1.8 K as shown in **Figure 2e**. This Hall resistivity signal can be related to the magnetization ($M$) of the sample through the following expression:[43–45]

$$\rho_{xy} = \frac{1}{ne}\mu_0 H + \mu_0 R_s M \qquad (2)$$

where $n$ is the carrier density, $e$ is the elementary charge, $\mu_0$ is the vacuum permeability, $H$ is the applied field, and $R_s$ is the anomalous Hall coefficient. The first term of Equation (2), $\mu_0 H/ne$, is referred to as the ordinary Hall effect (OHE), while the second term, $\mu_0 R_s M$, is the anomalous Hall resistivity, $\rho_{AHE}$. Analysis of the former (See SI for details) uncovers that the major charge carriers in this 2D crystal are holes, with a carrier density of $2.0 \times 10^{21}$ cm$^{-3}$ ($2.0 \times 10^{27}$ m$^{-3}$) at 1.8 K. These results, which are in agreement with measurements of bulk crystals (SI Figure 16), reveal the presence of hole pockets at the Fermi level after intercalation.[13,41,43] Although the OHE reveals valuable information about the basic electronic structure of our materials, it does not convey any information on magnetism. Insight into the long-range magnetic ordering of few-layer $Fe_xTaS_2$ was obtained by extracting the magnetization-dependent $\rho_{AHE}$ as well as measuring the magnetoresistance (MR), which reflects the evolution of scattering interactions between charge carriers and magnetic moments. The hysteretic behavior of $\rho_{AHE}$ and MR displayed in **Figure 2f** reveal that the material is a hard ferromagnet with a very large coercive field ($H_c$) that reaches 2.0 ± 0.1 T at 1.8 K. To the best of our knowledge, this is the largest $H_c$ value recorded among 2D ferromagnetic crystals to-date.[42,46–50]

The broadness of $\rho_{AHE}$ and MR switching profiles is consistent with the presence of inhomogeneous MEIs arising from some magneto-structural disorder. This behavior is reminiscent of off-stoichiometric bulk $Fe_xTaS_2$, in which crystallographic defects (Fe vacancies or interstitials) weaken the coupling between neighboring Fe centers.[20,31] The resulting inhomogeneous coupling of magnetic moments across the crystal broadens the hysteresis curves and may engender MR through spin-disorder scattering.[20] Indeed, we find that the maximum value of MR decreases as the temperature approaches the Curie temperature ($T_C$), which is estimated to be about 31 K from the inflection in $\rho_{xx}(T)$ shown in **Figure 2d**. This observed change in MR with temperature is consistent with a spin-disorder scattering mechanism.[20] Interestingly, while we might expect fully closed hysteresis loops above this ordering temperature, **Figure 2f** instead shows that both MR and $\rho_{AHE}$ traces retain their hysteretic behavior up to 40 K, consistent with the persistence of some long-range magnetic order above 31 K. This discrepancy in the apparent $T_C$ prompted us to further examine the magnetic ordering temperature.

### Magnetic phase transitions and magnetocrystalline anisotropy (MCA) in thin $Fe_xTaS_2$

By fitting the temperature dependence of the remanent $\rho_{AHE}$ signal to the power-law form $\alpha(1-T/T_C)^{\beta}$, it is possible to extract both the $T_C$ and the critical exponent ($\beta$) for the magnetic material (see SI for details).[42] $\beta$ provides insight into the nature of the magnetic phase transition in the crystal and enables us to identify its universality class, which characterizes the dimensionality of our system, the symmetry of its magnetization, and the range of spin–spin interactions.[51,52] This analysis and the discontinuity in $\rho_{AHE}(T)$ around 31 K before full hysteresis closure (**Figure 3a**), suggest the presence of two magnetic phase transitions at about 31 K and 41 K, which we discuss in detail later. The critical exponent of the low-temperature magnetic phase, $\beta = 0.31 \pm 0.04$, closely matches the theoretical value for the 3D Ising model $\beta = 0.325$,[53] which describes systems with uniaxial MCA. To experimentally examine the MCA, we compare the MR curves with the external field applied parallel to the *c*-axis ($H \parallel c$) with those in which the external field is applied along the *ab*-plane ($H \parallel ab$) in **Figure 3b**. The MR profile is hysteretic only for $H \parallel c$, consistent with an easy-axis along *c*, and a strong out-of-plane MCA in few-layer $Fe_xTaS_2$.

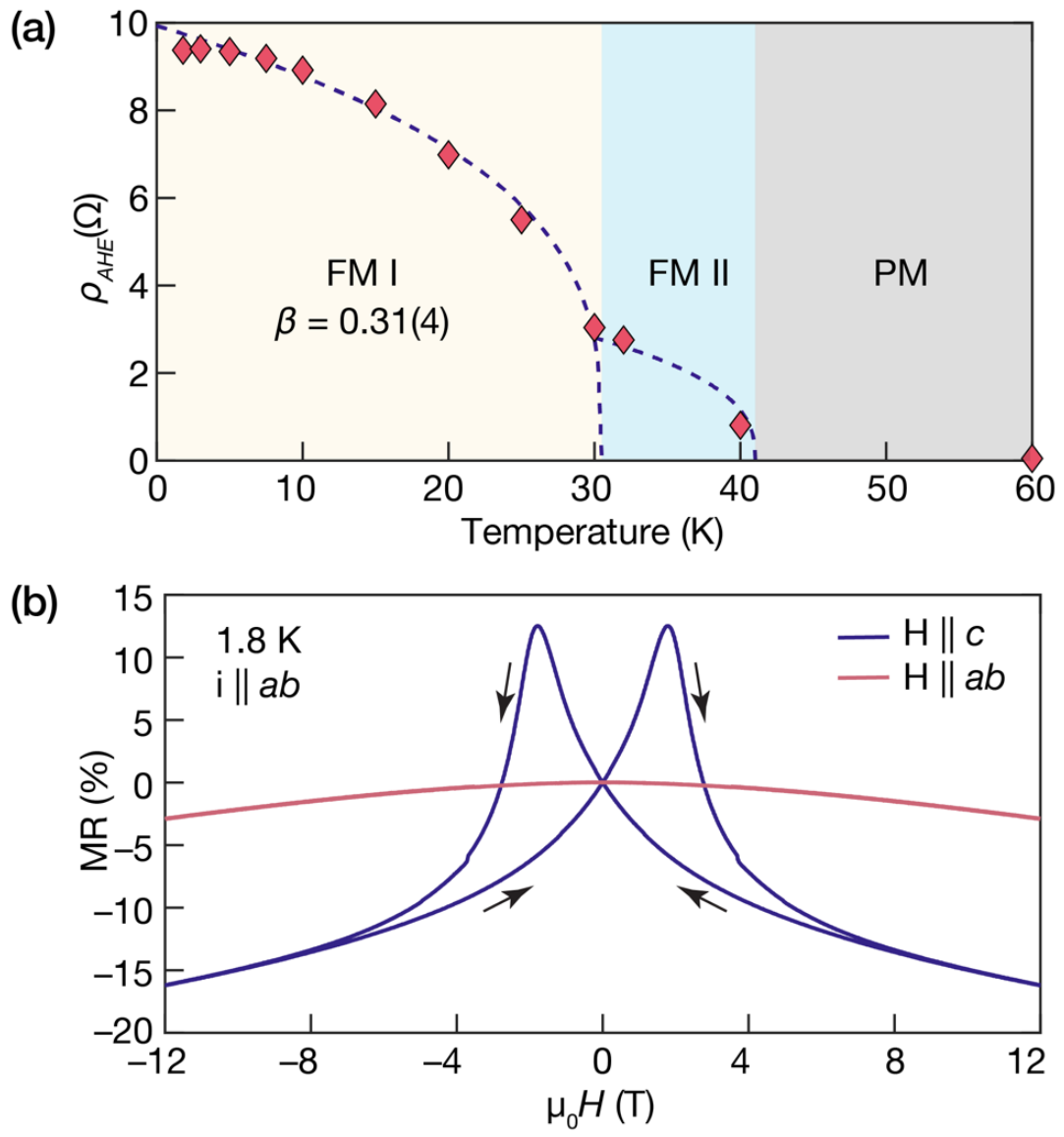

**Figure 3. (a)** Temperature dependence of the remanent $\rho_{AHE}$ for the 11-layer device of $Fe_xTaS_2$. Ferromagnetic (FM) and paramagnetic (PM) phases are highlighted. Dashed lines represent the least squares fit to the formula $\alpha(1-T/T_C)^{\beta}$. **(b)** MR of the device with the external field parallel and perpendicular to the $c$-axis.

**Magnetic behavior, transport, and Fe disorder in the ultrathin limit of $Fe_xTaS_2$.**

The large MCA in $Fe_xTaS_2$ prepared by chemical intercalation raised the possibility of anisotropy-stabilized long-range ferromagnetism in ultrathin $Fe_xTaS_2$ with even fewer layers. Hysteretic magnetotransport features of intercalated 2-, 3-, and 5-layer $TaS_2$ flakes (see SI Figure 4) confirm that hard ferromagnetism is retained down to the thinnest limit of $Fe_xTaS_2$ (**Figure 4a**). All measured devices (**Figure 4b**) exhibit hysteretic MR profiles only for $H \parallel c$, confirming their large out-of-plane MCA (SI Figure 21). This observation of ferromagnetism and large MCA in few-layer Fe intercalated $TaS_2$ is supported by Density Functional Theory (DFT) calculations (see SI for details). In these computations, we considered four $\sqrt{3}a \times \sqrt{3}a$ superlattices of $Fe_xTaS_2$ with $L$ = 2, 3, 4, 5 $TaS_2$ layers (see SI Figure 23). For all $L$-layer systems we find a high-spin ferromagnetic order with calculated Fe magnetic moments of 3.3–3.5 $\mu_B$ per site. For all calculated heterostructures and magnetic configurations, we find that Fe converges to a $d^6$ manifold (*i.e.* an $Fe^{2+}$ oxidation state), which aligns with literature on bulk $Fe_xTaS_2$. Consistent with our experiments, we find the easy-axis to be out-of-plane with a calculated MCA in the 5-layer case of 6 meV per formula unit (1.5 meV/Fe). The origin of this anisotropy is the large out-of-plane orbital moment, which was previously calculated to be of 1.0 $\mu_B$/Fe in the bulk case.[13] Notably, the large MCA in our crystals is expected to dominate over shape anisotropy, which is an important factor for thin films with smaller MCA.[54]

Additional magnetotransport measurements also reveal that these thin crystals have high carrier densities and hole carriers (SI Figure 16). However, despite these similarities, their magnetic behaviors are substantially different. The 5-layer $Fe_{0.38(2)}TaS_2$ crystal displays only slightly broadened $\rho_{xy}$ and MR hysteresis curves, suggesting a small degree of inhomogeneous MEIs in this material. In contrast, the 3-layer sample, $Fe_{0.37(2)}TaS_2$, which possesses a comparable stoichiometry to the 5L crystal, exhibits substantially broader magnetization switching. These data indicate that stoichiometry may not be the primary determinant of magnetic order in 2D $Fe_xTaS_2$. Instead, dimensionality may play the dominant role in shaping magnetic properties. Indeed, the $\rho_{xy}$ and MR features found in measurements of the 2-layer crystal, $Fe_{0.58(14)}TaS_2$ are comparable to those of the 3-layer crystal, despite a significant difference in stoichiometry. This strong influence of dimensionality on magnetic order may stem from its impact on intercalant ordering via electronic and/or structural effects. For example, measurements of bulk crystals of a related system, $Fe_xTiS_2$, reveal that the ordering of intercalants in the plane is sensitive to the structure of Fe intercalants along the $c$-axis.[55–57] By extension, we suggest that decreasing numbers of layers could result in a reduced tendency for ordering Fe centers in-plane. Moreover, in the low-dimensional limit, intercalation may produce distinct Fe intercalant structural phases that may not be present in bulk crystals, a behavior that can also be theoretically understood from the Mermin–Wagner framework as applied to 2D crystallization processes.[58]

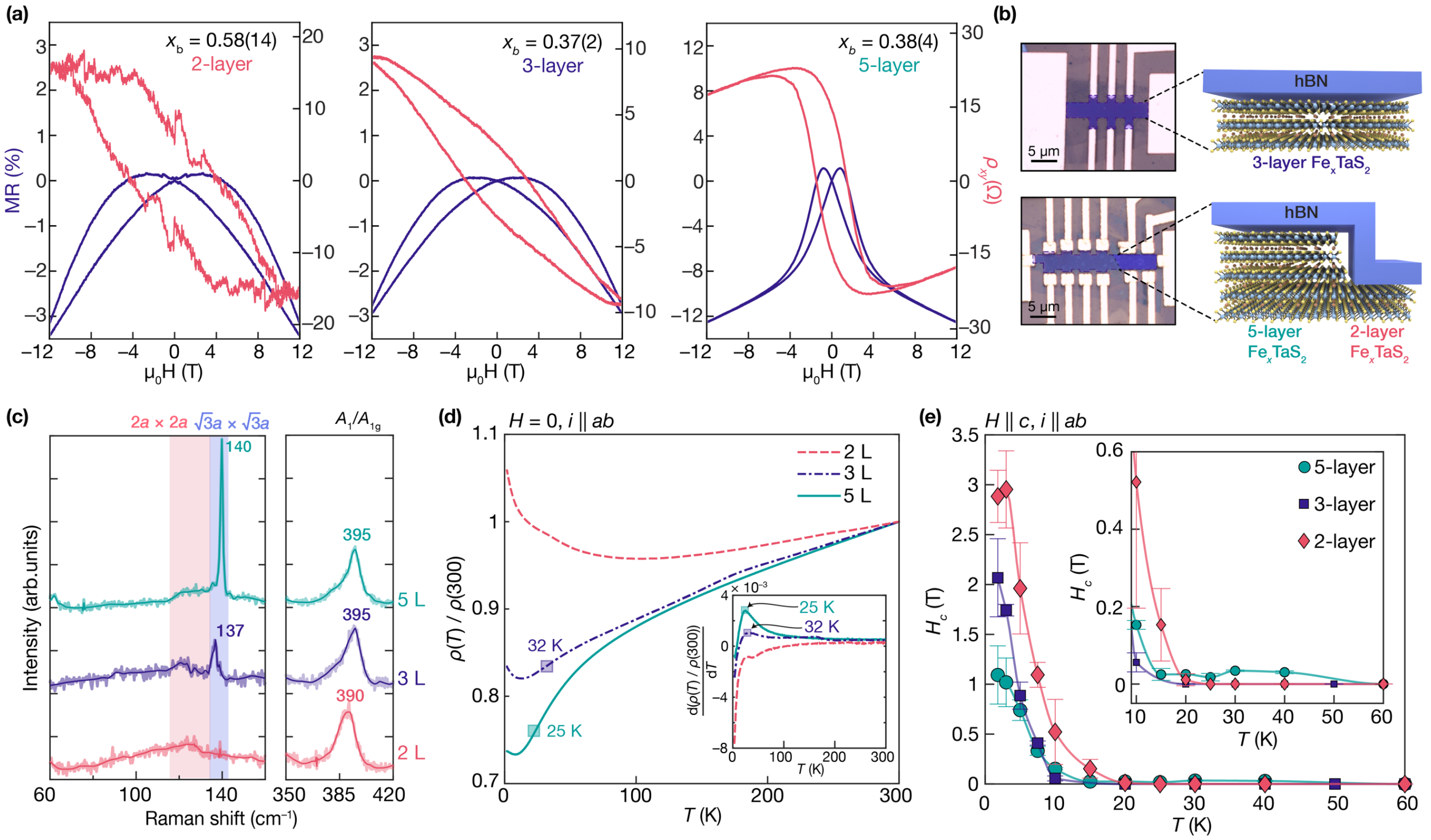

**Figure 4. (a)** Magnetotransport measurements, **(b)** optical micrographs with heterostructure schematics, **(c)** cumulative Raman spectra, **(d)** temperature-dependent longitudinal resistivities and their first derivatives with respect to temperature and **(e)** temperature-dependent coercivities for the 5-, 3- and 2-layer devices. In **(c)**, the Raman data is normalized to the $A_1/A_{1g}$ peak. In **(d)**, $(\rho(T)/\rho(300))/dT$ maxima are marked on the $\rho(T)/\rho(300)$ curves for the 3-layer and 5-layer devices.

To evaluate the extent of this intercalant ordering in these ultrathin intercalation compounds, we used confocal Raman spectroscopy to probe the Fe superlattice modes in the intercalated 5-, 3- and 2-layer crystals. **Figure 4c** shows that the superlattice structures and order in these few-layer devices indeed differ appreciably. We find that the 5-layer sample is mostly ordered in the $\sqrt{3}a \times \sqrt{3}a$ superstructure, with a hallmark sharp and intense Fe* feature close to 140 $cm^{-1}$.[32] In the 3-layer device, the characteristic $\sqrt{3}a \times \sqrt{3}a$ and $2a \times 2a$ superstructure modes are broader and closer in intensity, consistent with increased disorder and the presence of both superlattices, notwithstanding an effectively identical composition to the 5-layer crystal. Finally, the bilayer sample exhibits a substantially broad feature in the low-frequency region, pointing to a high level of intercalant disorder. Although sharp Fe-related modes are absent in this sample, the red-shifted $A_1$ mode reveals that the $TaS_2$ lattice has been electron-doped[36,37] and STEM-EDS confirms the substantial intercalation of Fe. Taken together, we suggest that structural disorder, which is encouraged by reduced dimensionality, likely promotes the observed broadened magnetic responses in ultrathin $Fe_xTaS_2$.

Traditionally considered an undesirable property of crystals, intercalant disorder plays a key, useful role in defining and manipulating the magnetic properties of magnetic intercalation compounds. In $Fe_xNbS_2$ crystals with $x$ near 1/3, disorder results in a spin-glass phase that coexists with the antiferromagnetic state, and this disorder-induced spin glass enables the switching of antiferromagnetic domain orientations and electronic conductivity with small electrical pulses.[59–63] In bulk $Fe_xTaS_2$ crystals, disorder has been demonstrated to increase ferromagnetic hardness, as defects can pin magnetic domains and increase the coercivity.[64] Congruently, apparent intercalant disorder in our bilayer and 3-layer $Fe_xTaS_2$ crystals may, together with intrinsically large MCA, be one origin for the observed giant coercivities reaching 2.9 ± 0.3 T and 2.1 ± 0.4 T, respectively (**Figure 4a**). In contrast, the coercivity of the more ordered 5-layer $Fe_xTaS_2$ is lower, amounting to 1.1 ± 0.3 T at 1.8 K.

Notwithstanding the very large coercivities measured for these few-layer magnetic crystals, it is noteworthy that these coercive fields are slightly lower than those measured for analogous bulk $Fe_xTaS_2$.[31]

These discrepancies in $H_c$ could suggest that the overall strength of MEIs may decrease in the few-layer limit. However, our analysis of ordering temperatures in these few-layer crystals reveals that the overall MEIs, while perhaps slightly weaker, may be quite comparable to those found in bulk crystals of similar composition. This insight was gleaned from a consideration of their Curie temperatures, determined from the temperature dependence of their longitudinal resistivities (**Figure 4d**) and coercivities (**Figure 4e**). Based on the inflections in $\rho_{xx}$ and $d\rho_{xx}/dT$ shown in **Figure 4d**, the 5- and 3- layer crystals magnetically order below 25 K and 32 K, respectively. However, as we discuss in more detail below, the closure of their hysteresis, *i.e.*, onset of $H_c = 0$, does not coincide with the inflections in $\rho_{xx}$ (**Figure 4e,** SI Figure 18,19). For the 5-layer crystal, the magnetic hysteresis instead closes between 40 K and 60 K, about 20 K higher than the temperature of inflection in $\rho_{xx}$. The hysteresis observed in the 3-layer crystal instead closes below the $\rho_{xx}$ inflection temperature (between 10 and 20 K), as observed in bulk $Fe_{1/3}TaS_2$.[41]

For the bilayer crystal, **Figure 4d** also shows that decreasing temperatures do not produce the characteristic sharp drop in $\rho_{xx}$ below a magnetic ordering temperature, as observed in crystals with $L \geq 3$ layers. Instead, $\rho_{xx}$ exhibits an upturn in resistivity for T < 100 K to reach a low-temperature resistivity greater than the $\rho_{xx}$ value measured at room temperature. This upturn in $\rho_{xx}$ has not been observed in bulk crystals of $Fe_xTaS_2$ and may be a result of Kondo-like scattering or weak localization[52,65–68] (SI Figure 22, SI Section 4.6), both of which can result in an increase in resistivity with decreasing temperature in metals.[52] Because $\rho_{xx}(T)$ for the bilayer sample does not exhibit any features typically attributed to a magnetic phase transition, we estimate $T_C$ as between 20 and 25 K based on the disappearance of hysteresis in $\rho_{xx}$ over this temperature range (**Figure 4e**, SI Figure 17). In summary, these measured ordering temperatures for few-layer $Fe_xTaS_2$ are comparable to the $T_C$ for $x = 0.33$ bulk crystals (around 38 K),[31,41] signifying that it is likely the intraplanar MEIs, which are retained in the 2D limit, that are the dominant coupling interactions in $Fe_xTaS_2$ flakes.

**Co-existing magnetic phases in few-layer $Fe_xTaS_2$**

As described above, in all measured few-layer samples, there exist discrepancies between ordering temperatures determined from inflections in $\rho_{xx}(T)$ and those measured from the closure of magnetic hysteresis. Moreover, the temperature dependence of remanent $\rho_{AHE}$ for the 5-layer $Fe_xTaS_2$ (SI Figure 20) appears to display more than one magnetic transition. This behavior, which was also observed for the 11-layer crystal (**Figure 3a**), again points to the presence of multiple co-existing ferromagnetic phases. Although one possible origin is superlattice inhomogeneity/disorder, these crystals display Raman spectra (**Figures 2b, 4c**) that are consistent with an overwhelmingly dominant $\sqrt{3}a \times \sqrt{3}a$ Fe superlattice. We therefore interrogated other possible origins of these magnetic transitions using DFT computations. Interestingly, in addition to the high-spin ground state, we also find metastable zero-spin and intermediate-spin solutions in our DFT calculations. For the 2-layer case, we find the zero-spin state is 461 meV per formula higher in energy than the high-spin (3.5 $\mu_B$/Fe) case. We also find an intermediate spin state with 2.1 $\mu_B$/Fe that is 273 meV per formula unit higher in energy than the high-spin case. In both cases, the magnetic moment reduction occurs with a decrease in the van-der-Waals gap—the interplanar S–S distance changes from 2.89 Å (high spin) to 2.83 Å (intermediate spin) to 2.56 Å (zero spin), with a corresponding reduction in Fe–S bond lengths. Along with this reduction in Fe moment and interlayer separation in going from the high- to intermediate-spin cases, we find that the Ta projected moment significantly increases from 0.03 $\mu_B$ per Ta in the high-spin case to 0.15 $\mu_B$ per Ta in the intermediate-spin case for those two Ta closest to the Fe. These Ta moments are antiferromagnetically coupled to the Fe moments. Our calculations suggest that the Ta acquires this net moment because of charge redistribution to the neighboring Ta sites with the reduction of the interlayer distance. We observe similar trends for the full $L$-layer series studied with changes in Fe magnetic moments resulting in modification of the interlayer separation and charge transfer to the neighboring Ta sites. We note, however, that in all cases we find a high-spin ground state.

These theoretical results point to other possible origins of the multiple magnetic phases observed in experiments, independent of intercalant/superlattice disorder. Firstly, we identify metastable spin states in our DFT calculations that are strongly dependent on the interlayer separation and the resulting Fe local environment. This strong structural dependence on the total magnetization could give rise to spin state transitions upon thermal lattice contraction. Secondly, we find that the Ta atoms adjacent to our Fe intercalants acquire a non-negligible moment (0.15 $\mu_B$/Ta in the 2-layer case) depending on the interlayer separation. This moment on Ta could also result in a separate magnetic ordering temperature. Future experimental and theoretical studies on 2D $Fe_xTaS_2$ spanning a greater range of stoichiometries, and degrees of intercalant ordering may reveal additional details on how magneto-structural disorder and lowered dimensionality affect the value of $T_C$ and competing long-range magnetic orders.

## CONCLUSIONS

We have demonstrated that long-range ferromagnetic order persists down to the thinnest limit of the magnetic intercalation compound $Fe_xTaS_2$. Combined structural, spectroscopic, and magnetoelectronic characterization indicates that both intercalant order/disorder and stoichiometry influence the magnetic properties of these low-dimensional materials. Nevertheless, overall, the high MCA expected from the coordination environment imposed by the vdW interface between $TaS_2$ sheets appears to stabilize long-range, hard ferromagnetic ordering in samples of all thicknesses. More broadly, our results demonstrate a versatile platform for developing 2D magnets with magnetic properties that can be manipulated by modulating their intercalation amount, dimensionality, and structural homogeneity. This work opens the door to modulating the MCA and MEIs of intercalated TMDs by marrying the synthetic versatility of soft chemical intercalation[23–27] with distinctive techniques developed for fabricating vdW architectures and controlling the emergent physics in atomically layered materials, such as layer count,[2,69] deterministic assembly of vdW heterostructures,[70,71] electrostatic gating,[4,5,72] strain[73,74] and the introduction of an interlayer twist.[75,76] Exploring this constellation of tuning knobs is poised to produce novel 2D intercalation compounds that enable the exploration of exotic emergent magnetic phases as well as materials for low energy switching of magnetism.

## ASSOCIATED CONTENT

### Supporting information

The Supporting information is available free of charge on the ACS Publications Website at: https://pubs.acs.org/doi/suppl/10.1021/jacs.2c02885/suppl_file/ja2c02885_si_001.pdf

**Notes**

The authors declare no competing financial interest.

**ACKNOWLEDGEMENTS**

The authors thank Oscar Gonzalez, Zhizhi Kong, Lilia Xie, and Steven Zeltmann for helpful discussions. This material is based upon work supported by the Air Force Office of Scientific Research under AFOSR Award No. FA9550-20-1-0007. We would like to thank Gatan, Inc. as well as P Denes, A Minor, J Ciston, C Ophus, J Joseph, and Ian Johnson at LBNL who contributed to the development of the 4D Camera used for STEM-DPC measurements. This research used resources of the National Energy Research Scientific Computing Center (NERSC), a U.S. Department of Energy Office of Science User Facility located at Lawrence Berkeley National Laboratory, operated under Contract No. DE-AC02-05CH11231 using NERSC award ERCAP0020898 and ERCAP0020897. M.V.W. acknowledges support from a UCB Chancellor's Fellowship and NSF Graduate Research Fellowship. Instrumentation used in this work was supported by grants from the W.M. Keck Foundation (Award # 993922), the Canadian Institute for Advanced Research (CIFAR–Azrieli Global Scholar, Award # GS21-011), the Gordon and Betty Moore Foundation EPiQS Initiative (Award #10637), and the 3M Foundation through the 3M Non-Tenured Faculty Award (#67507585). N.P.K. acknowledges support from the Hertz Fellowship and from the National Science Foundation Graduate Research Fellowship under Grant No. DGE-1745301. Confocal Raman spectroscopy was supported by a DURIP grant through the Office of Naval Research under Award No. N00014-20-1-2599 (D.K.B.). Experimental and theoretical work at the Molecular Foundry, LBNL was supported by the Office of Science, Office of Basic Energy Sciences, of the U.S. Department of Energy under Contract No. DE-AC02-05CH11231. Computational works were carried out using supercomputing resources of the National Energy Research Scientific Computing Center (NERSC) and the TMF clusters managed by the High Performance Computing Services Group, at LBNL. K.W. and T.T. acknowledge support from the Elemental Strategy Initiative conducted by the MEXT, Japan, Grant Number JPMXP0112101001, JSPS KAKENHI Grant Number JP20H00354 and the CREST (JPMJCR15F3), JST.

**ABBREVIATIONS**

MCA, magnetocrystalline anisotropy; OAM, orbital angular momentum; MEI, magnetic exchange interaction